# Evolving Diploid Boolean and Multi-Valued Gene Networks

Larry Bull

Computer Science Research Centre

University of the West of England, Bristol UK

Larry.Bull@uwe.ac.uk

**Abstract**

Boolean networks have been widely used to explore aspects of gene regulation, traditionally with a single network. A modified form of the model to explore the effects of increasing the number of gene states has also recently been introduced. In this paper, these discrete dynamical networks are evolved as diploids within rugged fitness landscapes to explore their behaviour. Results suggest the general properties of haploid networks in similar circumstances remain for diploids. The previously proposed inherent fitness landscape smoothing properties of eukaryotic sex are shown to be exhibited in these dynamical systems, as is their propensity to change in size based upon the characteristics of the network and fitness landscape.

**Introduction**

Random Boolean networks (RBN)[Kauffman, 1969] have been previously used in conjunction with the NK model [Kauffman & Levin, 1987] to explore the evolution of gene regulation within tuneable fitness landscapes [Bull, 2012]. The standard NK model assumes a binary gene alphabet but a recent extension to higher alphabets suggests the basic properties of the original model remains [Bull, 2022]. The effects of altering the size of the alphabet of the underlying gene expression state representation and logic in RBN have also recently been explored using the non-binary NK model [Bull, 2024] (after [Solé et al., 2000]). Again, results suggest that a number of the basic properties of the original binary model remain, whilst aspects such as how fitness is sampled and how many genes contribute explicitly to the fitness calculation can significantly vary behaviour. This paper explores the evolution of Boolean and multi-valued diploid gene regulatory networks (GRN) finding similar evolutionary behaviour to the equivalent haploid case both in terms of sensitivity to internal connectivity and the ruggedness of the fitness landscape. Diploids are shown to be more likely to increase in size compared to haploids, except when there are a large number of genes contributing directly to fitness.

**Random Multi-Valued Networks**

Within the traditional form of RBN, a network of $R$ nodes, each with $B$ directed connections randomly assigned from other nodes in the network, all update synchronously based upon the current state of those $B$ nodes. As the name suggests, gene states are traditionally from a binary alphabet ($A$=2) and use a randomly assigned Boolean update function. Hence those $B$ nodes are seen to have a regulatory effect upon the given node, specified by the Boolean function attributed to it. Since they have a finite number of possible states and they are deterministic, such networks eventually fall into an attractor. It is well-established that the value of $B$ affects the emergent behaviour of RBN wherein attractors typically contain an increasing number of states with increasing $B$ (see [Kauffman, 1993] for an overview). Three regimes of behaviour exist: ordered when $B$=1, with attractors consisting of one or a few states; chaotic when $B$>2, with a very large number of states per attractor; and, a critical regime around $B$=2, where similar states lie on trajectories that tend to neither diverge nor converge (see [Derrida & Pomeau, 1986] for formal analysis). Note that the size of an RBN is traditionally labelled $N$, as opposed to $R$ here, and the degree of node connectivity labelled $K$, as opposed to $B$ here. The change is adopted due to the traditional use of the labels $N$ and $K$ in the NK model of fitness landscapes which are also used in this paper, as will be shown.

As noted above, multi-valued logic forms of the original Boolean model have been explored. Following [Bull, 2024], in the simplest case, each node can exist in one of $A$ ($A≥2$) states and is assigned a randomly created logic table for each of the $A^B$ possible configurations (Figure 1). Figure 2 shows the typical number of nodes changing state per update cycle in such discrete dynamical systems where $R$=50, with various connectivity $B$ and number of gene expression states $A$, using 0<$B$<6 and 1<$A$<9. As can be seen, in these random multi-valued networks (RMN), for high connectivity ($B$>2) behaviour is significantly changed with increasing $A$. That is, significantly more nodes change state per update cycle when $A$>2 with such connectivity. Formal analysis suggests the critical regime occurs at $B$=1 with increasing $A$ [Solé et al., 2000].

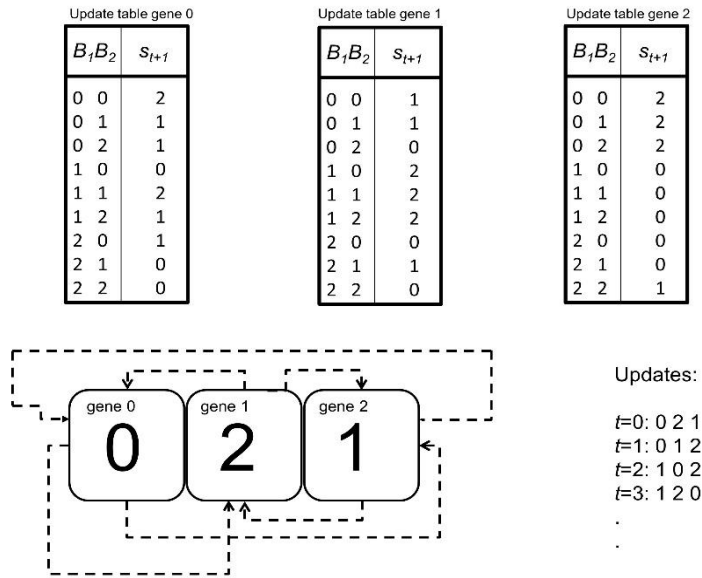

Figure 1. An example random multi-valued regulatory network model, with

$R$=3, $B$=2 and $A$=3.

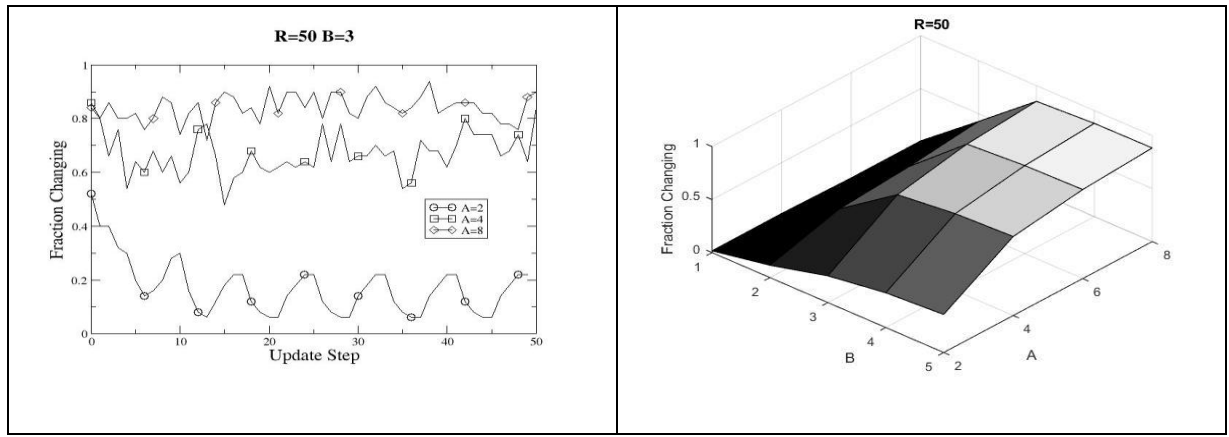

Figure 2. Showing the effects on the typical behaviour of the multi-valued regulatory networks with varying connectivity $B$ and states $A$ after 50 update cycles. Results are the average of one hundred randomly created networks per parameter configuration.

**The NK Model**

Kauffman and Levin [1987] introduced the NK model to allow the systematic study of various aspects of fitness landscapes. In the standard model, the features of the fitness landscapes are specified by two parameters: *N*, the length of the genome; and *K*, the number of genes that has an effect on the fitness contribution of each binary gene (*A*=2). Thus increasing *K* with respect to *N* increases the epistatic linkage, increasing the ruggedness of the fitness landscape. The increase in epistasis increases the number of optima, increases the steepness of their sides, and decreases their correlation [Kauffman, 1993]. The model assumes all intragenome interactions are so complex that it is only appropriate to assign random values to their effects on fitness. Therefore for each of the possible *K* interactions a table of $A^{(K+1)}$ fitnesses is created for each gene with all entries in the range 0.0 to 1.0, such that there is one fitness for each combination of traits. The fitness contribution of each gene is found from its table. These fitnesses are then summed and normalized by *N* to give the selective fitness of the total genome (Figure 3).

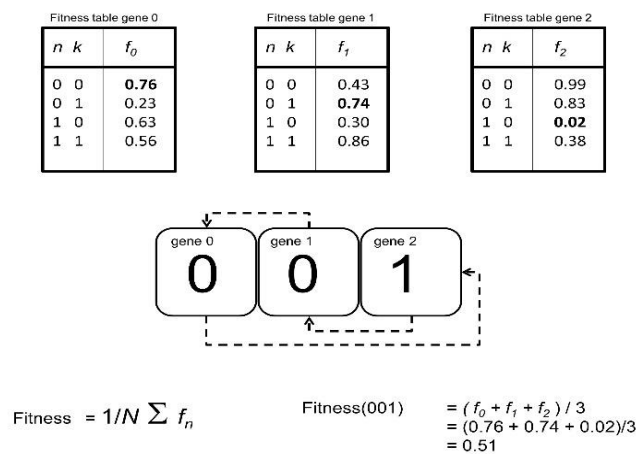

Figure 3. An example traditional binary NK model, with *N*=3 and *K*=1.

The traditional binary NK model has recently been extended to higher alphabets, i.e., fitness tables of size $A^{(K+1)}$ are created per gene, finding that the general properties of the landscapes are seemingly preserved [Bull, 2022]. This form of the NK model is here used to explore the evolutionary behaviour of the multi-valued diploid regulatory networks introduced above – an extended version of the RBNK model [Bull, 2012].

**The RMNK Model**

The combination of the discrete dynamical networks and NK model enables the exploration of the relationship between phenotypic traits and the genetic regulatory network by which they are produced [Bull, 2012]. In this paper, the following simple scheme is adopted: $N$ phenotypic traits are attributed to the first $N$ nodes within the network of $R$ genes (where $0<N\leq R$, Figure 4). Thereafter all aspects of the two models remain as described above, with simulated evolution used to evolve the RMN on NK landscapes. Hence the NK element creates a tuneable component to the overall RMN's fitness landscape.

**Evolving RMN**

Simulated evolution has previously been used to design haploid/single RBN, beginning with a simple feedforward network architecture [Van den Broeck & Kawai, 1990] (see [Bull, 2012] for an overview). Following [Kauffman, 1993], a mutation-based hill-climbing algorithm is used here, where the single point in the fitness space is said to represent a converged species, to examine the properties and evolutionary dynamics of the models. That is, the population is of size one and a mutation can either alter the logic function of a randomly chosen node or alter a randomly chosen connection for that node (equal probability).

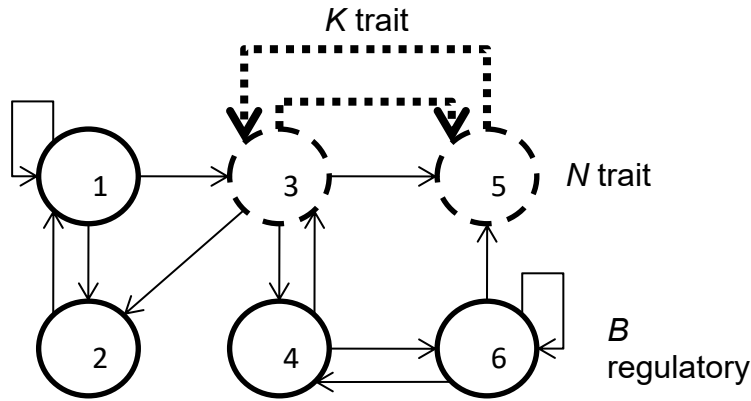

*R*=6, *B*=2, *N*=2, *K*=1

Figure 4. Example RMNK model for haploid networks. A network consists of *R* nodes, each node containing *B* integers in the range [1, *R*] to indicate input connections and an *A*-ary string of length $A^B$ to indicate the multi-valued logic function over those connections.

A single fitness evaluation of a given RMN is ascertained by first assigning each node to a randomly chosen start state (uniform in *A*) and updating each node synchronously for *U* update cycles. Here *U* is chosen such that the networks have typically reached an attractor (*U*=50). At update cycle *U*, the value of each of the *N* trait nodes is then used to calculate fitness on the given NK landscape. This process is repeated ten times on the given NK landscape, repeated for ten randomly created NK landscapes, i.e., 10x10=100 runs, with the fitness assigned to the RMN being the average fitness. The "population" is said to move to the genetic configuration of the mutated individual if its fitness is greater than the fitness of the current individual; the rate of supply of mutants is seen as slow compared to the actions of selection. Ties are broken at random.

Figure 5 shows the typical evolutionary performance of $R$=50 RMN with various internal connectivity $B$ (0< $B$ <6) and logic alphabet $A$ (1< $A$ <9), on landscapes of varying ruggedness $K$ (0 and 4), after 5000 generations. As previously demonstrated in [Bull, 2024], when $N$=10, fitness generally decreases with increasing $B$, regardless of $K$ or $A$. That is, results for $B$=1 or $B$=2 are always statistically better (T-test, $p<0.05$) than for $B$=4 or $B$=5. When $K$=0, increasing $A$ typically decreases fitness regardless of $B$. The relative decrease in fitness is highest when $A$ >2 and $B$ >2, with $B$<3 RMN seemingly most robust to increasing $A$. When $K$>0 and $B$<3 fitnesses increase with increasing $A$. Fitnesses are all roughly equally low for $B$>2, regardless of $A$. As predicted [Solé et al., 2000], fitnesses are typically highest for $B$=1 and increasing $A$. Figure 6 shows the effects of increasing the number of nodes by which fitness is explicitly calculated, with $N$=$R$. As can be seen, the same general behaviour as for $N$=10 emerges. However, the drop in fitness for increasing $B$ from $B$=1 to $B$=2 is much larger and fitness levels are generally decreased for all $B$ and $A$, regardless of $K$ (T-test $p<0.05$ comparing each $N$=10 with $N$=$R$ cases). That is, it appears to be a significantly more difficult task, perhaps as might be expected.

In the above, fitness is calculated from the state of the $N$ trait nodes on the step after $U$ network update cycles, i.e., typically within an attractor. To explicitly consider the evolution of temporal behaviour, i.e., particular sequences of gene activity, the state of the RMN can be sampled on every update cycle, i.e., up to and including within an attractor. Here total fitness is calculated as the average of the fitness of each successive state of the $N$ nodes for $U$ cycles. Thus, networks must evolve temporal behaviour which keeps them consistently within the high optima region(s) of the fitness landscape. Figure 7 shows examples of how the change causes a significant decrease in fitness (T-test, $p<0.05$) achieved with any $K$ and $N$ for $B$>2 and $A$>2. Fitnesses are not significantly affected otherwise (T-test, $p≥0.05$). Figure 2 showed how the $A$=2 RMN, i.e., traditional RBN, experienced fewer numbers of nodes changing state for higher $B$ compared to higher $A$. Thus, it appears the evolutionary process finds it harder to produce consistently high fitness phenotypic behaviour with such networks under constant evaluation.

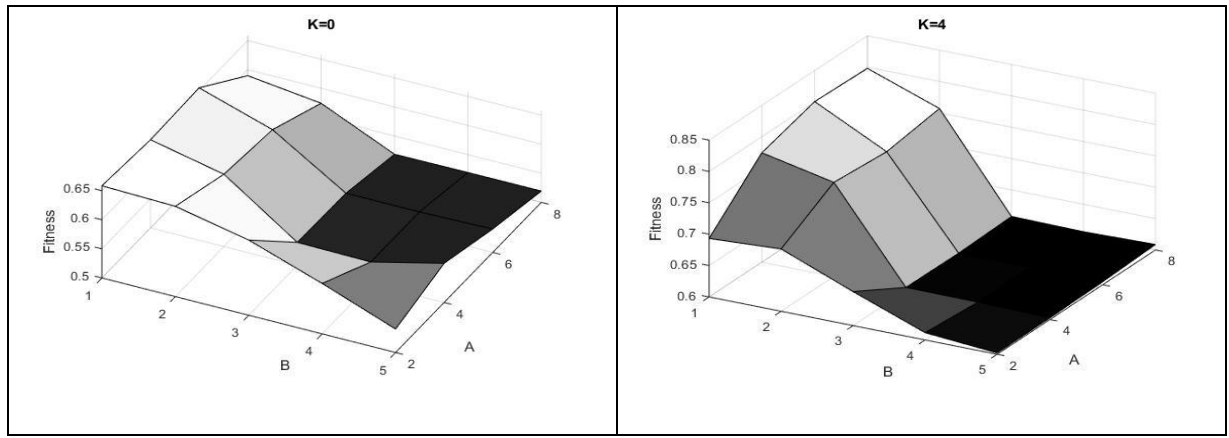

Figure 5. Showing fitness reached after 5000 generations for combinations of haploid network connectivity (*B*), different logic alphabets (*A*), for different degrees of fitness landscape ruggedness (*K*), with *R*=50 and *N*=10.

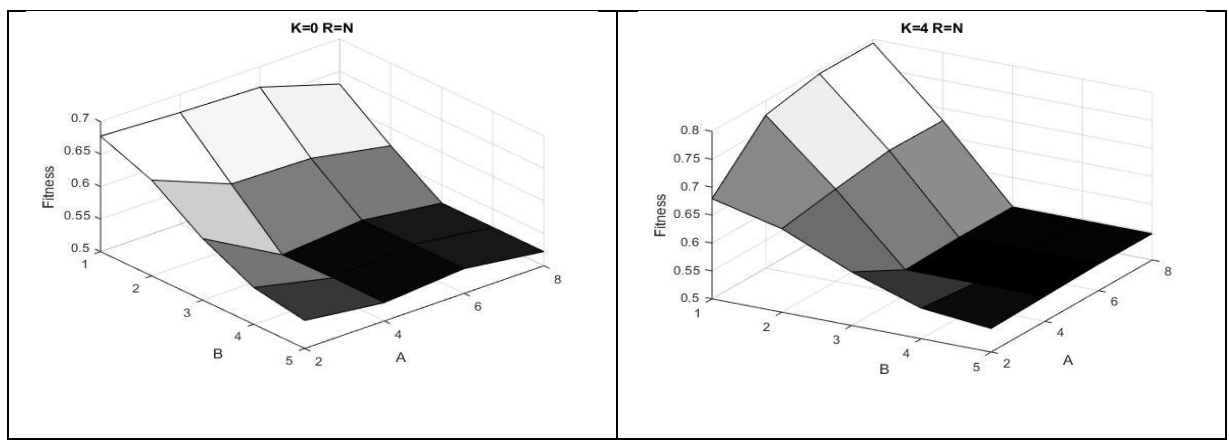

Figure 6. Showing the same configurations as in Figure 5 but with *R*=*N*.

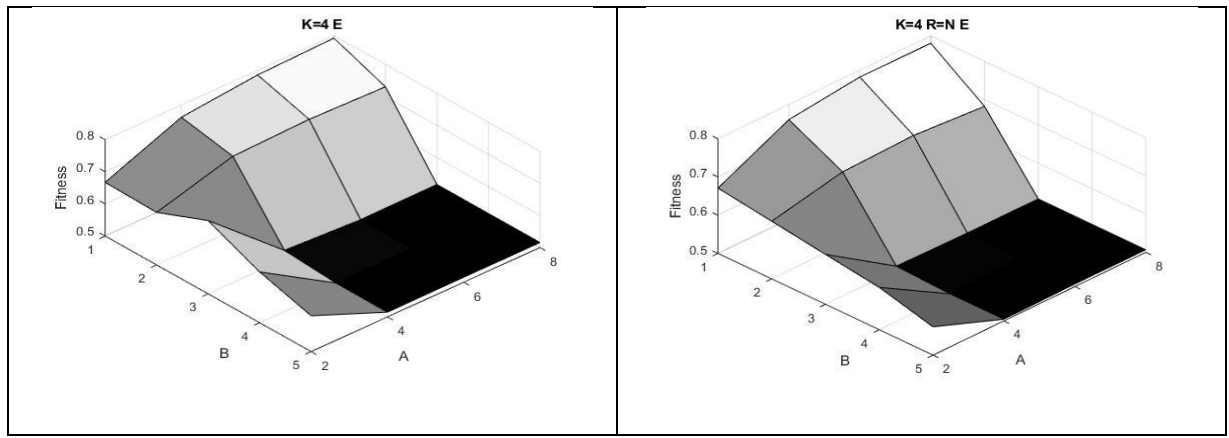

Figure 7. Showing examples of the *K*=4 configurations in Figures 5 and 6 but with fitness calculated as the average for the network state on each update step.

**Evolving Diploid RMN**

There appears to be no previous work on the evolution of diploid RBN other than briefly in [Bull, 2020, pp82-84] ([Emmerich et al., 2015] used a single network to represent a diploid). A continuous-valued diploid GRN model has recently been explored [Okubo & Kaneko, 2022] and ordinary difference equation models of simple diploid networks have also been presented (e.g., [Plahte et al., 2013]). A number of explanations have been suggested for why diploidy, or increasing ploidy in general, is beneficial, typically based around the potential for "hiding" mutations within extra copies of the genome (e.g., see [Otto, 2007] for an overview). Moreover, a change in ploidy can potentially alter gene expression, and hence the phenotype – through mutations, through epigenetic mechanisms, through rates of changes in gene product concentrations, no or partial or co-dominance, etc. That is, the fitness of the cell/organism is a combination of the fitness contributions of the composite haploid genomes.

Almost all eukaryotes reproduce sexually. The few explanations as to why a form of meiosis exists which includes a genome doubling stage — the diploid temporarily becomes a tetraploid — range from DNA repair (e.g., [Bernstein et al., 1988]) to the suppression of potentially selfish/damaging alleles (after [Haig & Grafen, 1991]). Explanations for the recombination stage vary from the removal of deleterious mutations (e.g., [Kondrashov, 1982] to avoiding parasites (after [Hamilton, 1980]) (see [Bernstein & Bernstein, 2010] for an overview). It has been suggested [Bull, 2017] that the emergence of sex – defined as successive rounds of syngamy and meiosis in a haploid-diploid lifecycle - enabled the exploitation of fitness landscape smoothing, i.e., a rudimentary form of the Baldwin effect [Baldwin, 1896]. Key to this explanation for the evolution of sex is to view the process from the perspective of the constituent haploids. A diploid organism may be seen to simultaneously represent two points in the underlying haploid fitness landscape. The fitness associated with those two haploids is therefore the fitness achieved in their combined form as a diploid; each haploid genome will have the same fitness in the diploid and that will almost certainly differ from that of their corresponding haploid organism due to the interactions between the two genomes. That is, the effects of haploid genome combination into a diploid can be seen as a simple form of phenotypic plasticity for the individual haploids before they revert to a solitary state during reproduction. In this way evolution can be seen to be both assigning a single fitness value to the region of the landscape between the two points represented by a diploid's constituent haploid genomes, i.e., a simple form of generalization, and altering – smoothing - the shape of the underlying haploid fitness landscape. The Baldwin effect is here defined as the existence of phenotypic plasticity that enables an organism to exhibit a significantly different (better) fitness than its genome directly represents. Over time, as evolution is guided towards such regions under selection, higher fitness alleles/genomes which rely less upon the phenotypic plasticity can be discovered and become assimilated into the population (after [Hinton & Nowlan, 1987]).

To extend the haploid model to diploid an individual simply contains two RMN, each run as described above for $U$ updates, and their individual fitnesses are averaged to determine the diploid's fitness. The reproduction process is altered to approximate two-step meiosis: each RMN is copied under mutation as above, non-sisters are recombined using a single randomly chosen point and one of the four resulting RMN is chosen at random. The process is repeated twice to form a new diploid (Figure 8).

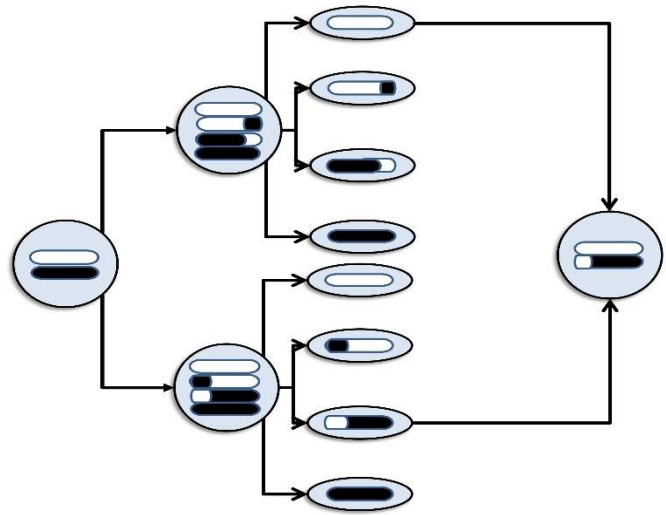

Figure 8. The two-step meiosis process as implemented with a converged population.

Figure 9 shows the typical evolutionary performance of $R$=50 diploid RMN with various internal connectivity $B$ (0<$B$<6) and logic alphabet $A$ (1<$A$<9), on landscapes of varying ruggedness $K$, after 5000 generations, with fitness calculated after $U$ updates. As with the haploids, fitnesses reached with $B$<3 are always highest, regardless of $K$ and $A$. Figure 10 shows examples of the same general behaviour seen in the haploids when fitness is calculated on every step, with fitnesses decreasing with increasing $A$, when $B$>2.

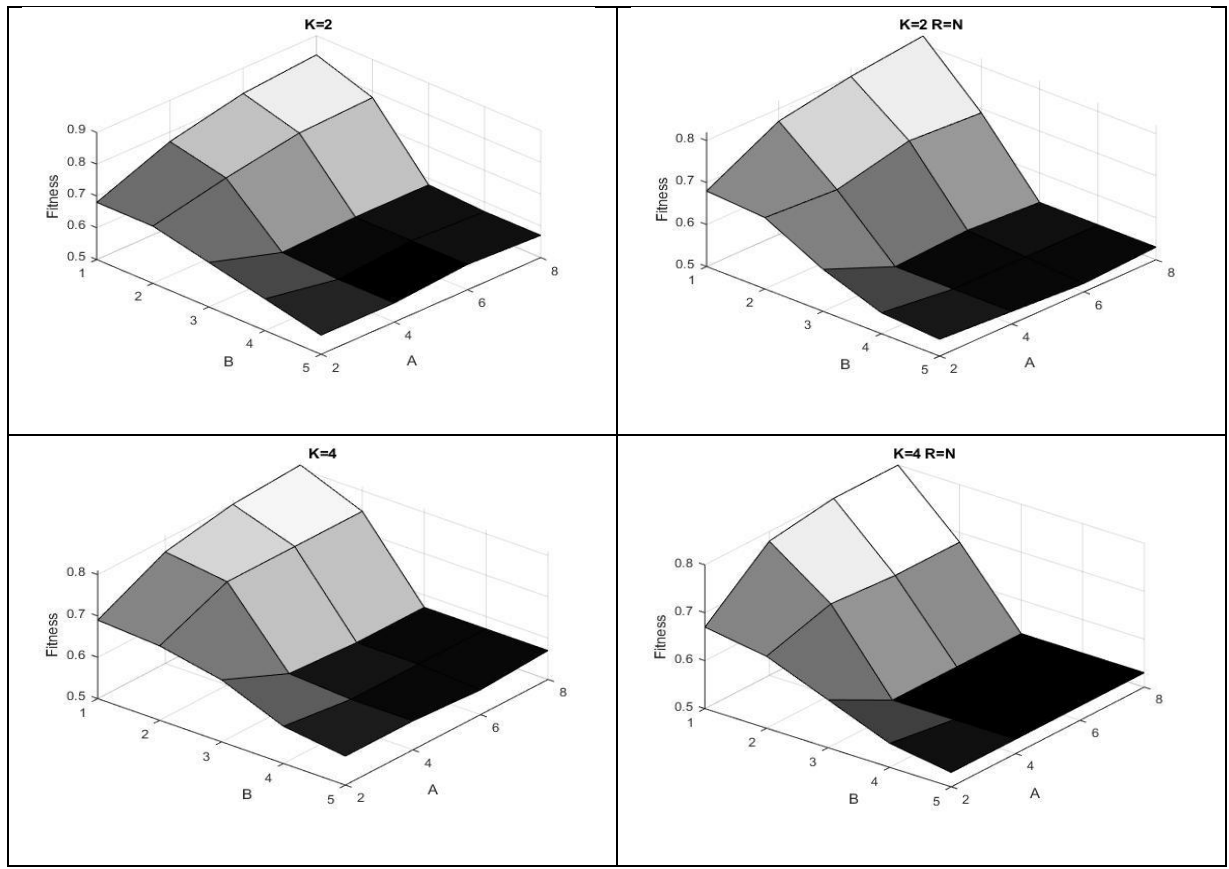

Figure 9. Showing fitness reached after 5000 generations for combinations of diploid network connectivity (*B*), different logic alphabets (*A*), for different degrees of fitness landscape ruggedness (*K*), with *R*=50.

.

There is no significant difference in the fitness reached by haploids and diploids for any combination of *B* and *A,* when *K*<6, regardless of *N*. The aforementioned work demonstrating the benefits of syngamy and meiosis used the original NK model (*A*=2), finding increased fitness compared to haploids for *K*>0 [Bull, 2017]. With *A*=2, fitnesses are seen to be highest for the diploid under both fitness approaches when *K*≥6 and *N*=10, in comparison to both the equivalent haploid and asexual diploid case (examples shown in Figure 11). The same is true for *N*=*R* when *K*≥8 (not shown). With higher *A*, the picture is less clear. It has previously been shown how the most beneficial amount of learning/phenotypic plasticity varies with the underlying ruggedness of the fitness landscape, as well as genome size, etc. [Bull, 1999]. Some amounts of learning were also shown to be detrimental for certain combinations. Figure 12 shows an example of an observed trend that diploids with higher $A$ and low $B$ become fitter than the equivalent haploids with higher $K$, although it has not been found to be universally true, as the result in [Bull, 1999] perhaps somewhat predicts. For example, with $A$=3 sex and diploidy is not significantly fitter than the equivalent asexual diploid or haploid RMN with $K$=6 but becomes fitter with $K$=8 and $B$<3, although it is worse for high $B$ in that case.

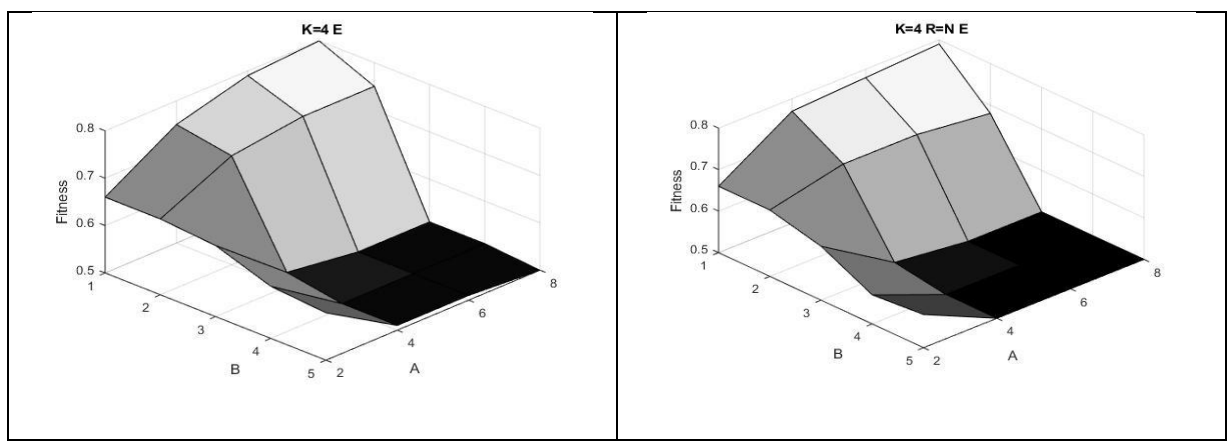

Figure 10. Showing examples of the $K$=4 diploid configurations in Figure 9 with fitness calculated as the average for the network state on each update step.

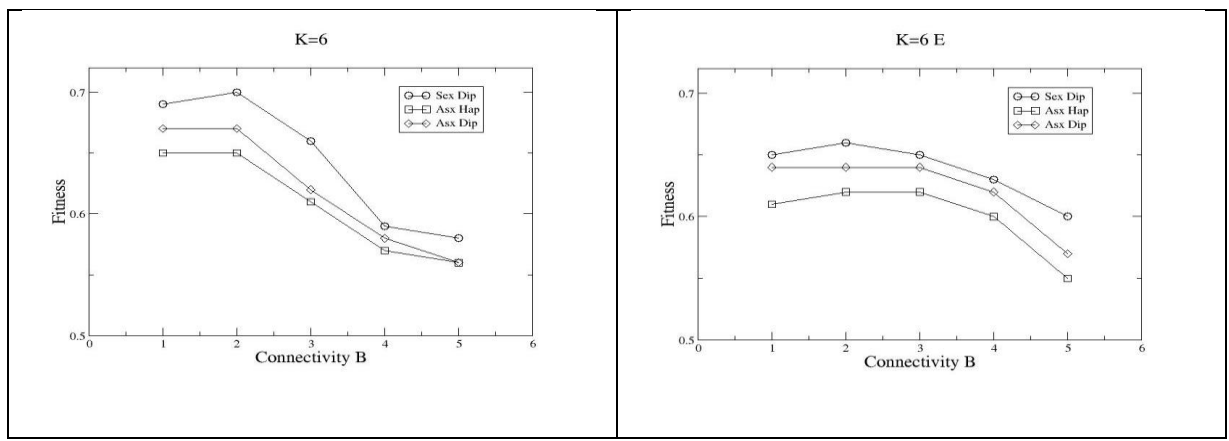

Figure 11. Showing comparative fitnesses of asexual haploid and diploid RBN.

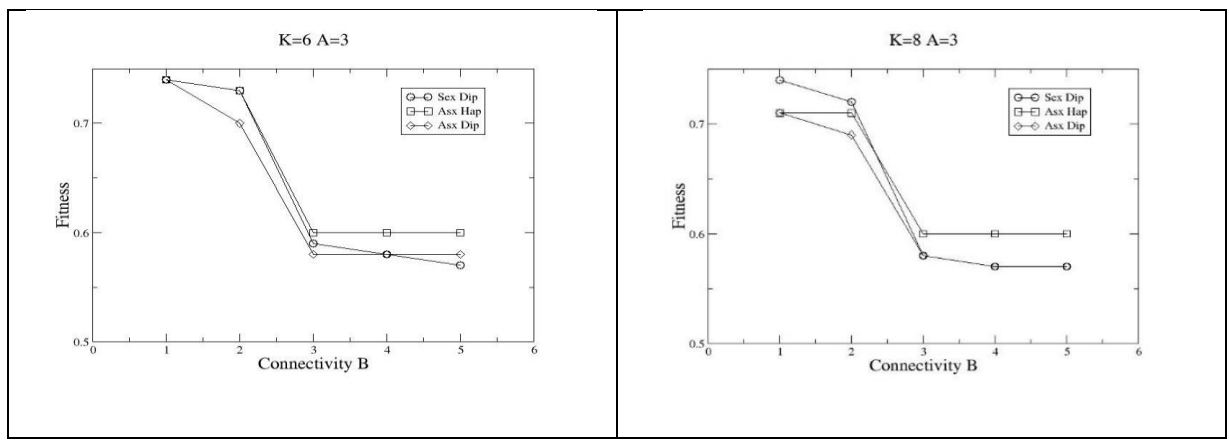

Figure 12. Showing comparative fitnesses of asexual haploid and diploid RMN with $A$=3.

**Genome Size**

Novel sequences of DNA can originate through a variety of mechanisms including retrotransposons, horizontal gene transfers, during recombination events, whole genome duplications, etc. For example, it is estimated that over half the genes in GRN are the result of gene duplications (e.g., [Teichmann & Babu, 2004]), a process that may aid robustness as well as providing a mechanism for subsequent innovation through function divergence (e.g., [Wagner, 2008]). Aldana et al. [2007] examined the effects of adding a new, single gene into a given RBN through duplication and divergence. They showed the addition of one gene typically only slightly alters the attractors of the resulting RBN when $B$<3 but that attractor structure is not conserved for higher $B$.

It has previously been suggested that increases in genome length are an inherent property of evolution on rugged fitness landscapes [Bull, 2020, pp.]. In [Bull, 2024], the experiments reported above with haploid RMN have been repeated with the addition of two extra "macro" mutation operators: one to delete a randomly chosen node (the $N$ trait nodes cannot be deleted), randomly re-assigning all of its connections; and, one to duplicate an existing node, connecting it to a randomly chosen node in the network. These two operators occur with equal probability to the two previously described mutation operators above, i.e., one of four mutations are chosen to create the offspring per generation. The replacement process is also altered such that, when fitnesses are equal, the smaller network is kept, with ties again broken at random. Networks are initialised at size $R$, as before, and labelled as of size $R'$ thereafter.

No significant change in the fitness of solutions is seen with the macro-structure mutation operators added regardless of whether $N=10$ or $N=R$ (not shown). However, as can be seen in Figure 13, when $N=10$, regardless of $K$, the networks decrease significantly in size when $B<3$ (T-test, $p<0.05$). The decrease in size increases with decreasing $A$. Moreover, $A=2$ networks decrease in size when $B<4$. That is, not only do low connectivity networks evolve the highest fitnesses for all $K$ and $A$, they are able to do so with a smaller number of nodes $R'$. It is known that both the number of states in an attractor and the number of attractors are dependent upon $R$ within traditional RBN, and that the general form of those relationships changes for low and high connectivity. For example, when $B=2$, attractors are typically of size $R^{0.5}$, whereas, when $B=R$, attractors typically contain $0.5 \times 2^{R/2}$ states (e.g., see [Kauffman, 1993] for a summary). Hence, regardless of $A$, the evolutionary process appears able to exploit the potential for ever smaller attractors for the low $B$ cases, driven by the additional selection pressure for network size reduction, and to do so whilst maintaining fitness. This result is somewhat anticipated by those of Aldana et al. [2007] but is in the opposite direction and with $A>2$: small reductional changes are maintained as the attractor space appears to be sufficiently conserved in both directions.

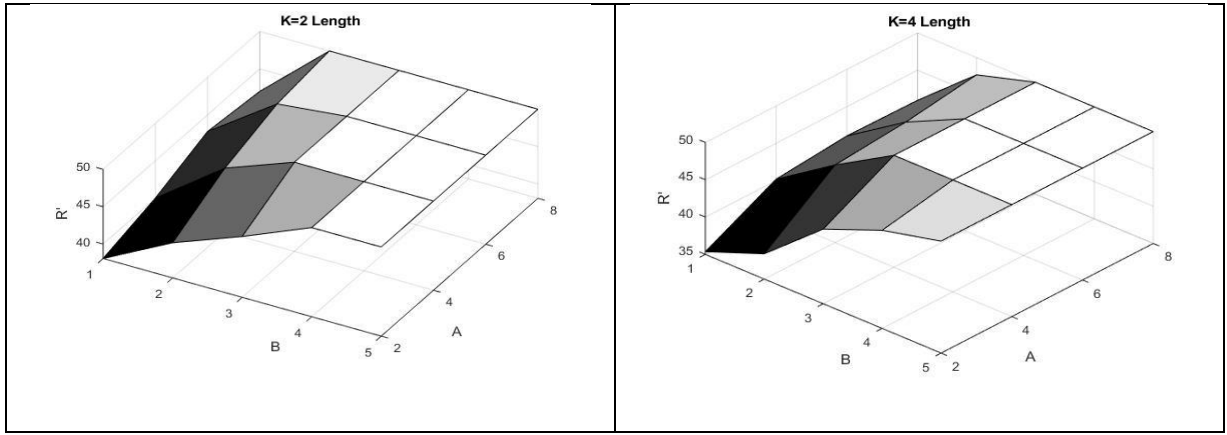

Figure 13. Showing lengths reached after 5000 generations for combinations of haploid network connectivity (*B*), different logic alphabets (*A*), for different degrees of fitness landscape ruggedness (*K*), with *R*=50 and *N*=10.

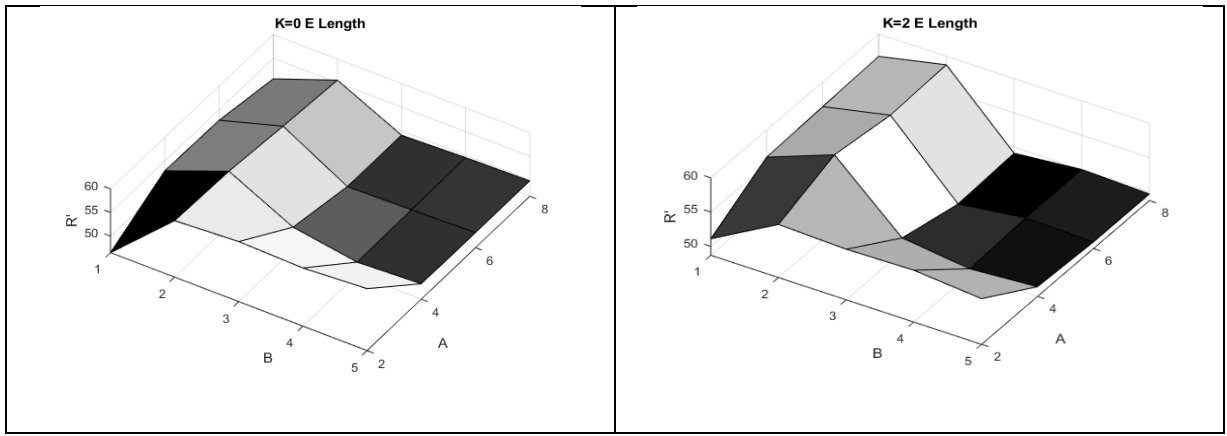

Figure 14. Showing the same haploid configurations as in Figure 13 with fitness calculated as the average for the network state on each update step.

Figure 14 shows examples of the effects on network size explicitly considering the evolution of temporal behaviour by sampling the state of the RMN on every update cycle in haploids. Again, there is no significant effect on fitness (not shown) but there is a change in the type of growth seen from the single point (attractor) fitness sampling case. Regardless of *N, A*, and

*K*, size is typically highest for *B*<3. However, when *A*=2, networks are largest with *B*>1. That is, significant growth typically occurs where the highest fitnesses emerge in such networks.

That networks do not decrease in size here for *N*=10 suggests that the removal of genes is more disruptive than the addition described by Aldana et al. [2007]: when the path through basins of attraction explicitly contributes to the overall fitness of the RMN, it seems gene deletion causes more change to the basins than addition. That is, gene deletion appears to affect the basins of attractors more than the attractors themselves since networks sampled after *U* updates experienced significant size reduction for low *B*.

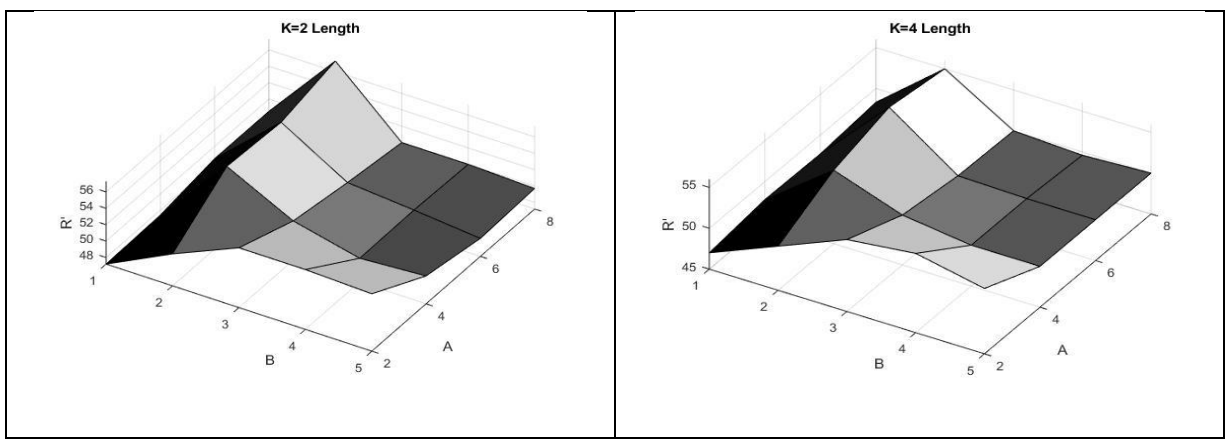

Figure 15. Showing lengths reached after 5000 generations for combinations of diploid network connectivity (*B*), different logic alphabets (*A*), for different degrees of fitness landscape ruggedness (*K*), with *R*=50 and *N*=10.

Figure 15 shows the same results for the diploid RMN introduced here. As with the haploids, networks decrease in size for *B*=1 for all *A*. However, they increase in size for *B*>2 with the highest growth seen for *B*=2 for *A*>2. That is, the results look akin to those seen in the haploids with the constant fitness applied. Fitnesses are not significantly different to without growth for *N*=10 (not shown) but are typically significantly worse for *N*=*R* (not shown) and quite why this should be the case is unclear. Figure 16 shows similar growth with the constant fitness applied, again fitnesses decrease when *N=R* (not shown). When growth proved beneficial, there was typically more growth for low *B* than in the haploid case. A similar result of increased growth in diploids has previously been reported in the traditional NK model [Bull, 2020, pp.40-43].

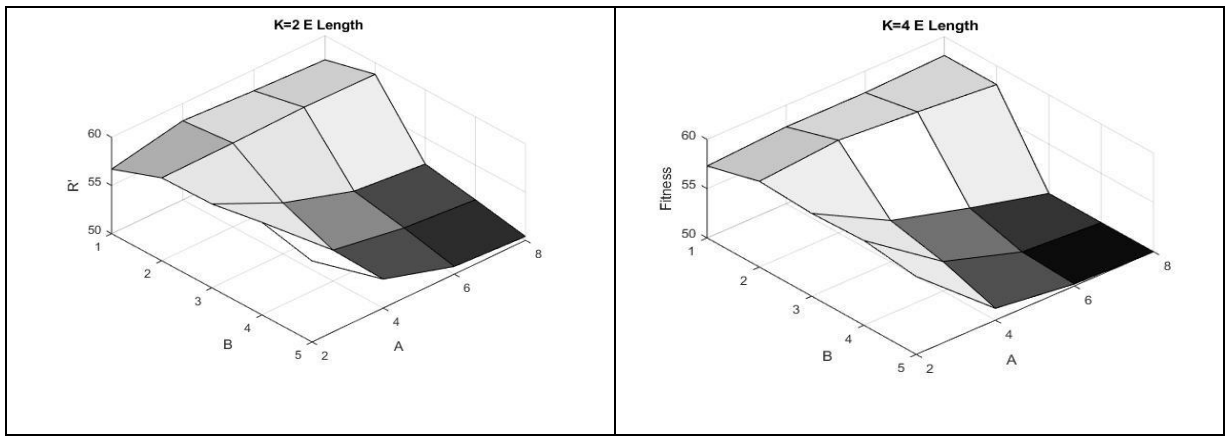

Figure 16. Showing the same configurations as in Figure 15 but with fitness calculated as the average for the network state on each update step.

**Conclusion**

Whilst Boolean models of haploid GRN have proven useful both theoretically and practically, e.g., in mammalian cells [Faur´e et al., 2006], Drosophila [Albert & Othmer, 2003], yeast [Li et al., 2004], amongst others, they clearly represent a simplification of the biology of many organisms of interest. This paper has introduced and explored evolving multi-valued diploid GRN under different conditions. Results suggest the many of the basic properties of the equivalent haploid systems are maintained but with some differences, particularly around genome growth.

Previous work on evolution of syngamy and meiosis using the traditional NK model suggested increasing benefits with increasing fitness landscape ruggedness [Bull, 2017]. The results here appear to concur for *A*=2 but are less clear with increasing *A*. It has been shown that the number of recombination points used can vary the benefits of meiosis [Bull, 2017] and that the related division of a genome in to chromosomes can also lead to further benefits from meiosis [Bull, 2020, pp.51-57]. Future work will consider the division of RMN in to chromosomes.